\documentclass[9pt, conference]{IEEEtran}
\IEEEoverridecommandlockouts

\usepackage{balance}
\usepackage{cite}
\usepackage{amsmath,amssymb,amsfonts}
\usepackage{algorithmic}
\usepackage{graphicx}
\usepackage{textcomp}
\usepackage{hyperref}
\usepackage{url}
\usepackage{multirow}
\usepackage{caption}
\usepackage{subcaption}
\usepackage{booktabs}
\usepackage{mathtools}
\usepackage{xcolor}
\usepackage{bm}

\def\BibTeX{{\rm B\kern-.05em{\sc i\kern-.025em b}\kern-.08em
    T\kern-.1667em\lower.7ex\hbox{E}\kern-.125emX}}
\begin{document}

\title{Text-Aware Adapter for Few-Shot Keyword Spotting}

\makeatletter
\newcommand{\linebreakand}{%
  \end{@IEEEauthorhalign}
  \hfill\mbox{}\par
  \mbox{}\hfill\begin{@IEEEauthorhalign}
}
\makeatother

\author{\IEEEauthorblockN{1\textsuperscript{st} Youngmoon Jung}
\IEEEauthorblockA{\textit{AI Solution Team} \\
\textit{Samsung Research}\\
Seoul, South Korea \\
youngm.jung@samsung.com}
\and
\IEEEauthorblockN{2\textsuperscript{nd} Jinyoung Lee}
\IEEEauthorblockA{\textit{AI Solution Team} \\
\textit{Samsung Research}\\
Seoul, South Korea \\
jin\_0.lee@samsung.com}
\and
\IEEEauthorblockN{3\textsuperscript{rd} Seungjin Lee}
\IEEEauthorblockA{\textit{AI Model Team} \\
\textit{Samsung Research}\\
Seoul, South Korea \\
sjsr.lee@samsung.com}
\linebreakand
\IEEEauthorblockN{4\textsuperscript{th} Myunghun Jung}
\IEEEauthorblockA{\textit{AI Solution Team} \\
\textit{Samsung Research}\\
Seoul, South Korea \\
mh95.jung@samsung.com}
\and
\IEEEauthorblockN{5\textsuperscript{th} Yong-Hyeok Lee}
\IEEEauthorblockA{\textit{AI Solution Team} \\
\textit{Samsung Research}\\
Seoul, South Korea \\
yong\_h.lee@samsung.com}
\and
\IEEEauthorblockN{6\textsuperscript{th} Hoon-Young Cho}
\IEEEauthorblockA{\textit{AI Solution Team} \\
\textit{Samsung Research}\\
Seoul, South Korea \\
h.y.cho@samsung.com}
}

\maketitle

\begin{abstract}
Recent advances in flexible keyword spotting (KWS) with text enrollment allow users to personalize keywords without uttering them during enrollment. 
However, there is still room for improvement in target keyword performance.
In this work, we propose a novel few-shot transfer learning method, called text-aware adapter (TA-adapter), designed to enhance a pre-trained flexible KWS model for specific keywords with limited speech samples.
To adapt the acoustic encoder, we leverage a jointly pre-trained text encoder to generate a text embedding that acts as a representative vector for the keyword. By fine-tuning only a small portion of the network while keeping the core components' weights intact, the TA-adapter proves highly efficient for few-shot KWS, enabling a seamless return to the original pre-trained model.
In our experiments, the TA-adapter demonstrated significant performance improvements across 35 distinct keywords from the Google Speech Commands V2 dataset, with only a 0.14\% increase in the total number of parameters.
\end{abstract}


\section{Introduction}
Keyword spotting (KWS) is a technique for detecting pre-defined keywords in audio streams. Unlike fixed KWS \cite{Chen14-ICASSP, Sainath-INTERSPEECH, TANG17-ICASSP}, which requires users to exclusively use specific keywords, flexible KWS allows users to utilize any custom keyword.
Custom keywords can be enrolled in flexible KWS either through audio \cite{Chen-ICASSP, Huang-ICASSP, Kurmi-INTERSPEECH} or text \cite{He-ICLR, Jung-INTERSPEECH-AdaMS, Nishu-INTERSPEECH, Lee-INTERSPEECH, RPL-Jung-INTERSPEECH}. 
Since text-based keyword enrollment does not require multiple utterances of a target keyword and can be achieved easily via text input, the demand for Text-enrolled Flexible KWS (TF-KWS) is growing.

TF-KWS systems typically use a text encoder for enrollment and an acoustic encoder for testing, both of which are optimized using deep metric learning (DML) \cite{Wang-CVPR} objectives such as contrastive loss \cite{Nishu-INTERSPEECH}, triplet-based loss \cite{He-ICLR}, and proxy-based loss \cite{Jung-INTERSPEECH-AdaMS, RPL-Jung-INTERSPEECH}.
As discussed in \cite{Wang-CVPR}, DML aims to learn an embedding space where the embedding vectors of similar samples are pulled closer together, while those of dissimilar samples are pushed apart.
Specifically for TF-KWS, the text embedding (TE) is learned to act as a representative vector for its corresponding keyword, thus attracting the acoustic embedding (AE) of the same keyword and repelling AEs of different keywords in the shared embedding space.

While TF-KWS models can support an unlimited number of keywords, their performance does not match that of keyword-specific models trained with abundant data for each keyword \cite{Bluche-INTERSPEECH}.
Thus, there remains potential to enhance performance for specific target keywords.
Given the challenge of collecting large amounts of data for a particular keyword, this problem can be approached through few-shot learning \cite{Kao-SLT}.
To address this, we propose a few-shot transfer learning approach called text-aware adapter (TA-adapter).
To the best of our knowledge, this is the first work on applying few-shot transfer learning to TF-KWS. 

Our research aims to adapt a small portion of the pre-trained acoustic encoder to a target keyword using limited speech data, leveraging the TE extracted from the corresponding text encoder. 
The TA-adapter consists of three main components: text-conditioned feature modulation (TCFM), feature weight adapter (FW-adapter), and TE classifier. 
Due to its modular design, the TA-adapter enables seamless restoration to the original pre-trained model, facilitating rapid adaption to various target keywords.
In our experiments, we evaluate performance of our method on the Google Speech Commands (GSC) V2 dataset  \cite{speechcommandsv2} under noisy and reverberant conditions.

\begin{figure*}[t]
  \centering
  \includegraphics[width=\textwidth]{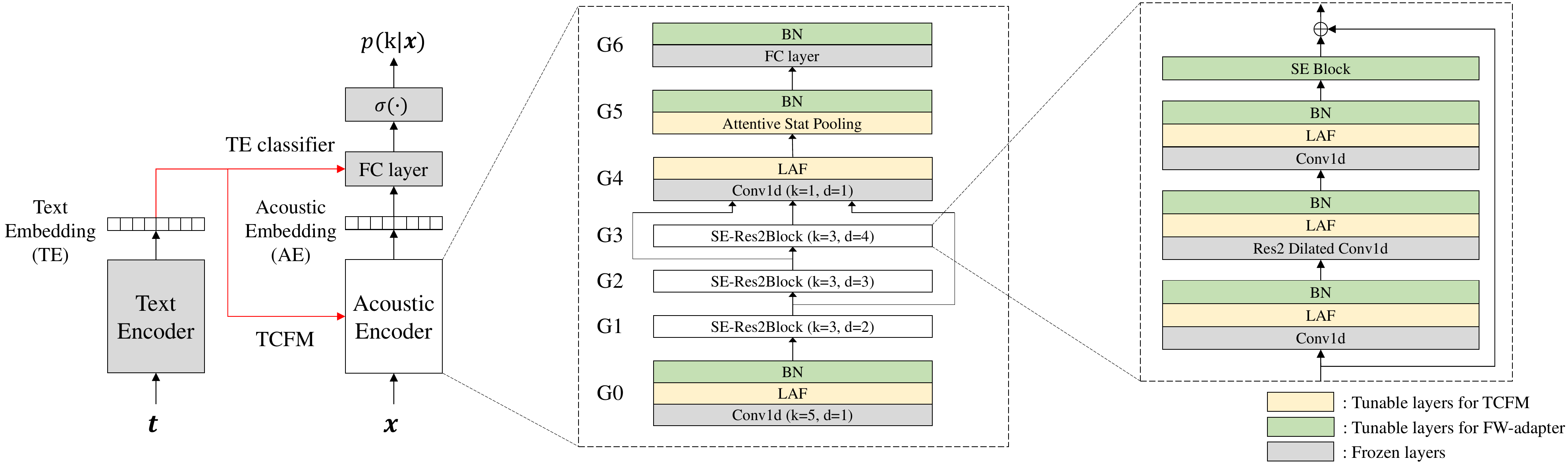}
  \vspace{-0.45cm}
  \caption{Overall architecture of text-aware adapter (TA-adapter). $\bm{t}$ and $\bm{x}$ represent input text and speech associated with keyword $k$. The red line indicates text embedding (TE) classifier and text-conditioned feature modulation (TCFM).}
  \label{fig:fig1}
  \vspace{-0.25cm}
\end{figure*}

\section{Related Work}
\label{sec:related_work}

Several studies have investigated few-shot transfer learning approaches for speech-enrolled flexible KWS \cite{Mazumder-INTERSPEECH, Awasthi-INTERSPEECH, Kao-SLT, Jung-metric-ICASSP, Yuan-INTERSPEECH}. 
In \cite{Kao-SLT}, the authors combined self-supervised learning (SSL) models with meta-learning algorithms. 
Many of these works insert an additional classification layer specific to target keywords following a pre-trained acoustic encoder, which is then trained either with or without freezing the pre-trained model.

The adapter approach was first introduced in computer vision and NLP \cite{Rebuffi-NIPS, Pfeiffer-EMNLP} to adapt large-scale SSL models for downstream tasks.
Adapter modules are inserted between intermediate layers of the pre-trained model and fine-tuned while keeping the rest of the model's parameters fixed.
Due to the minimal number of additional parameters within these adapter modules, the adapter approach prevents overfitting and has gained popularity for parameter-efficient fine-tuning of SSL models.
Inspired by this, our TA-adapter updates only a small subset of network parameters using few-shot samples, while maintaining the core acoustic encoder intact. 
Although the acoustic encoder in TF-KWS is not as large as SSL models, the adapter is well-suited for few-shot KWS due to its parameter efficiency and modular design.

In \cite{Navon-ICASSP}, the authors proposed AdaKWS, where a text encoder generates the parameters of Adaptive Instance Normalization (AdaIN) \cite{Huang-ICCV} layers in Keyword Adaptive Modules (KAMs), conditioning keyword information into an acoustic encoder.
However, unlike our approach, AdaKWS is not designed for few-shot KWS and requires a substantial amount of training data to jointly train both encoders from scratch. 
Consequently, it results in employing KAMs that contain excessive parameters for few-shot KWS.
To condition keyword information with fewer parameters more suitable for few-shot KWS, we adopt a learnable activation function (LAF) \cite{Ramos-ICLR} in our TCFM, while freezing the text encoder.
Conditioning by learning activations in \cite{Ramos-ICLR} showcased the effectiveness for conditioning speaker embeddings for personalized sound enhancement and speaker-dependent automatic speech recognition.

According to \cite{Sarfjoo-INTERSPEECH, Wang-INTERSPEECH}, squeeze-and-excitation (SE) \cite{Hu-CVPR} blocks and batch normalization (BN) layers are effective light-weight structures for low-resource domain adaptation in speaker recognition tasks. 
The underlying assumption is that essential speaker features have already been effectively learned by a pre-trained model using extensive training data. 
Therefore, for domain adaptation, it is sufficient to solely adjust how lower-level features are aggregated into higher-level features, which SE and BN handle by considering feature importance.
In our TA-adapter, we extend this concept to few-shot KWS through the FW-adapter.

\section{Text-aware adapter}
\label{sec:ta-adapter}

For the pre-trained model, we trained the acoustic and text encoders using Relational Proxy Loss (RPL) \cite{RPL-Jung-INTERSPEECH}, which exploits the structural relationships within AEs and TEs to promote their tight alignment.
We employed the ECAPA-TDNN architecture \cite{Desplanques-arxiv} as the acoustic encoder, which has been widely adopted in various speech processing tasks as an embedding extractor \cite{Zhao-ICASSP, Desplanques-arxiv, Liao-ICASSP, Li-INTERSPEECH, RPL-Jung-INTERSPEECH}.

The ECAPA-TDNN is built upon 1D convolutional layers and comprises multiple blocks similar to other deep CNN architectures, equipped with SE modules. 
As depicted in Fig. \ref{fig:fig1}, it consists of three SE-Res2Blocks, integrating the Res2Net \cite{Gao-TPAMI} with SE modules. In this architecture, the Res2Net processes multi-scale features extracted from various hierarchical levels, while the SE module performs channel-wise feature recalibration by learning feature weights based on their importance.
The output feature maps from all the SE-Res2Blocks are concatenated along the channel dimension, followed by a dense layer that processes the combined information to generate the features for attentive statistics pooling. The pooling layer converts variable-length features into a fixed-dimensional AE. 
In Fig. \ref{fig:fig1}, an LAF is applied for TCFM, which will be detailed in Section \ref{sec:tcfm}, replacing the ReLU used in the original ECAPA-TDNN.

\subsection{Text-conditioned feature modulation}
\label{sec:tcfm}
\begin{figure}[t]
  \centering
  \includegraphics[width=\columnwidth]{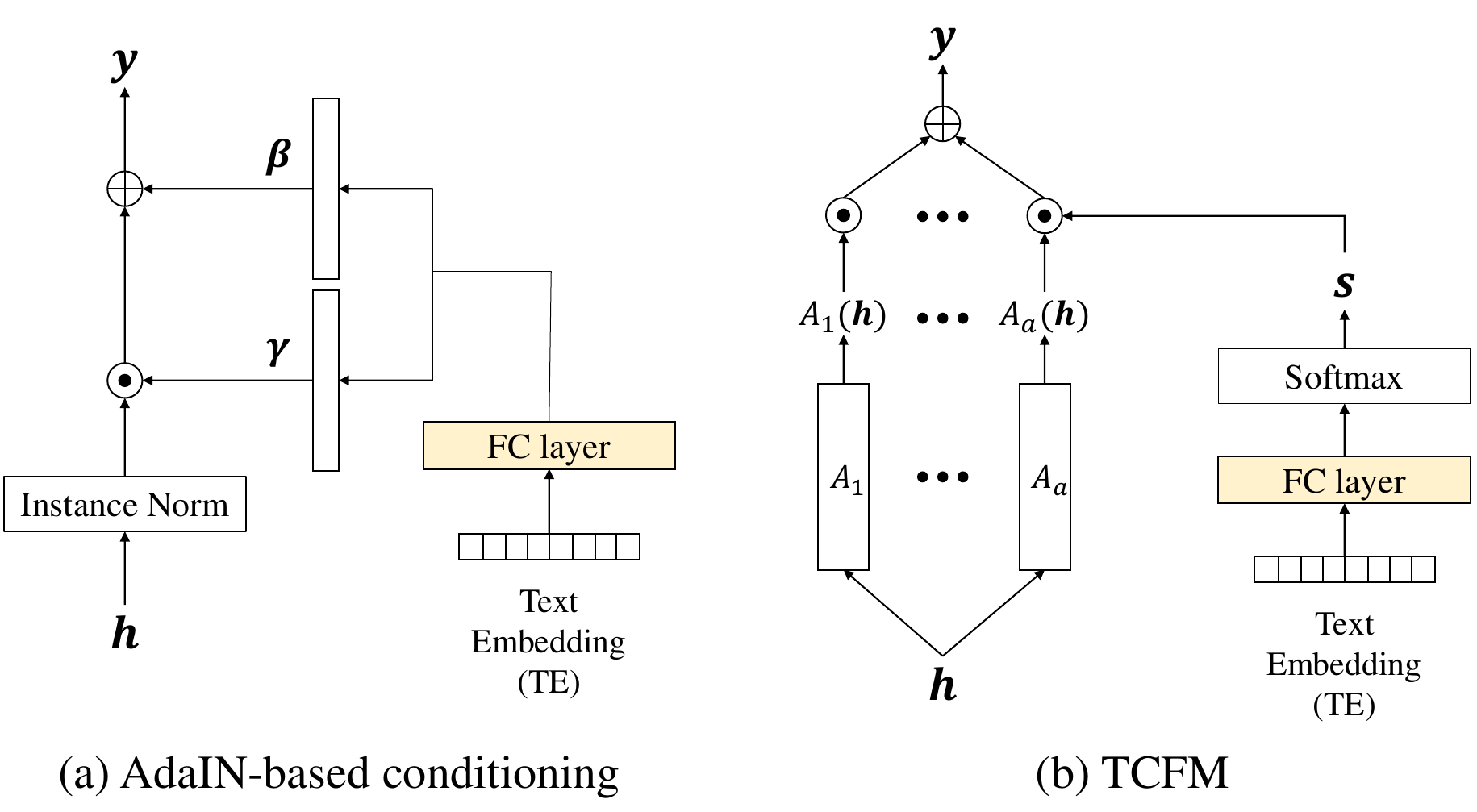}
  \vspace{-0.3cm}
  \caption{Comparison between (a) AdaIN-based conditioning and (b) text-conditioned feature modulation (TCFM).}
  \label{fig:fig2}
  \vspace{-0.1cm}
\end{figure}

First, for the TA-adapter, we adopt TCFM to transfer the target keyword information from the TE into the pre-trained acoustic encoder. 
By freezing the text encoder, we extract a TE for the target keyword, which serves as a representative vector for its corresponding keyword.
Fig. \ref{fig:fig2} highlights the difference between AdaIN-based conditioning and TCFM.
With AdaIN (Fig. \ref{fig:fig2}a), the scale and bias vectors $\{\bm{\gamma}, \bm{\beta}\}$ are estimated through a linear projection of the conditioning vector, TE, and applied to the instance-normalized features in a channel-wise manner.
In contrast, with TCFM (Fig. \ref{fig:fig2}b), we replace the ReLU of ECAPA-TDNN with LAF (see Fig. \ref{fig:fig1}).
TCFM conditions the keyword information from the TE by learning a weighted combination of basic activation functions based on the TE, requiring significantly fewer parameters.
Given an ordered set of $a$ basic activation functions $\{A_1, ..., A_a\}$, LAF is defined as follows:
\begin{equation}\label{eqn:eq1}
 \bm{s} = \textrm{softmax}(\textrm{TE}\cdot\bm{w} + \bm{b}),\,\bm{y}=\textrm{LAF}(\bm{h}|\textrm{TE})=\sum_{i=1}^a s_i A_i(\bm{h}),
\end{equation}
where $\textrm{TE} \in \mathbb{R}^{1 \times d}$, $\bm{w} \in \mathbb{R}^{d \times a}$, $\bm{b} \in \mathbb{R}^{1 \times a}$, and $\bm{s} \in \mathbb{R}^{1 \times a}$. 
For each LAF, unique trainable parameters $\bm{w}$ and $\bm{b}$ are optimized to estimate the activation weight vector $\bm{s}$ using the TE, thereby transforming the input features $\bm{h}$ into the activated features $\bm{y}$. 

The AdaIN-based conditioning requires $d \times 2 \times f$ parameters, where $f$ denotes the number of channels in $h$.
In comparison, our TCFM only requires $d \times a$ parameters.
Specifically, we set $d$, $f$, and $a$ to be 512, 256, and 6, respectively.
We employ six activation functions from the set of basic activation functions in \cite{Ramos-ICLR}, selected based on their validation performance: ELU, hard sigmoid, ReLU, softplus, swish, and tanh.
For more details, please refer to \cite{Ramos-ICLR}.

\subsection{Feature weight adapter}
\label{sec:fw_adapter}

In addition to feature modulation, we aim to refine the weighting and aggregation process of features within the acoustic encoder by adjusting attention weights and activation distributions.
We hypothesize that essential keyword features have already been effectively learned by the pre-trained acoustic encoder using extensive training samples.
Therefore, when transferring information about the target keyword with limited samples, it is sufficient to adjust the aggregation of low-level features into higher-level ones by emphasizing feature importance, which can be achieved through SE and BN.
As shown in Fig. \ref{fig:fig1}, we adapt only the BN and SE modules, highlighted in green.
Our experiments demonstrate that TCFM and the FW-adapter work synergistically, complementing each other.

\begin{table}[t]
\caption{GSC V2 dataset used in the experiment.}
\label{tab:tab1}
\footnotesize
\centering
\renewcommand{\arraystretch}{0.9}
\resizebox{\columnwidth}{!}{\begin{tabular}{c|c|c|cc|cc}
\hline
\multirow{2}{*}{Data} & \multirow{2}{*}{Keywords}  & \multirow{2}{*}{\# Train}                                                                                                                                                                                                                         & \multicolumn{2}{c|}{Valid}                  & \multicolumn{2}{c}{Test}                        \\ \cline{4-7} 
                          &    &                                                                                                                                                                                                                                                & \multicolumn{1}{c|}{\# Pos.} & \# Neg. & \multicolumn{1}{c|}{\# Pos.} & \# Neg. \\ \hline
Seen                      & \begin{tabular}[c]{@{}c@{}}`backward', `follow', \\ `forward', `happy', \\ `house', `one', \\ `seven', `sheila', \\ `visual', `zero'\end{tabular}                                                                                       & 2332               & \multicolumn{1}{c|}{231}            &      9750       & \multicolumn{1}{c|}{249}            & 10756            \\ \hline
Unseen                    & \begin{tabular}[c]{@{}c@{}}`bed', `bird', `cat',\\ `dog', `down', `eight',\\ `five', `four', `go',\\ `learn', `left', `marvin',\\ `nine', `no', `off',\\ `on', `right', `six',\\ `stop', `three', `tree',\\ `two', `up', `wow', `yes'\end{tabular} & 3090  &  \multicolumn{1}{c|}{307}           & 9674            & \multicolumn{1}{c|}{341}            & 10664            \\ \hline
\end{tabular}}
\vspace{-0.1cm}
\end{table}

\subsection{Text embedding classifier}
\label{sec:te_classifier}

The TE classifier process is illustrated by the red line in Fig. \ref{fig:fig1} (labeled as `TE classifier').
When TCFM and FW-adapter are applied simultaneously, an AE is extracted and then passed through the final fully-connected (FC) layer with weights $\theta_k \in \mathbb{R}^{d \times 1}$, generating output logits for the target keyword $k$. 
A sigmoid activation function is then applied to the output, producing $p(k|\bm{x})$, which represents the probability that the input speech $\bm{x}$ belongs to keyword $k$.
The acoustic encoder is adapted using binary cross-entropy loss.
Instead of learning a new weight vector from scratch, we fix the TE as the weight vector in the final classification layer.
This is reasonable because the TE has already been trained as a representative vector for its corresponding keyword during the pre-training phase using DML.
Thus, we define the inner product between the fine-tunable AE and the fixed TE as the logits for $k$.
Since both AE and TE are L2-normalized, this operation is equivalent to calculating their cosine similarity, $S(\cdot,\cdot)$.
Therefore, the final score can be obtained in the same manner as in TF-KWS.
We can express $p(k|\bm{x})$ as follows:
\begin{equation}\label{eqn:eq2}
 p(k|\bm{x}) = \sigma(\theta_k \cdot \textrm{AE}) = \sigma(S(\textrm{TE}, \textrm{AE})).
\end{equation}

\section{Experiments}
\label{sec:typestyle}

\subsection{Experimental setup}
\label{sec:typestyle}

The pre-training strategy followed the same approach as our previous work in \cite{RPL-Jung-INTERSPEECH}, including the use of identical datasets, acoustic features, and model architectures. 
For the TA-adapter, we used the GSC V2 dataset \cite{speechcommandsv2} containing 35 keywords. 
Compared to the training set used for pre-training \cite{Speechocean-DB}, 10 of these keywords were seen during pre-training, while 25 were unseen (as shown in Table \ref{tab:tab1}).
We evaluated the model under three low-resource scenarios: 5-shot, 10-shot, and 15-shot learning. 
For each keyword, we developed three separate models by randomly selecting 5, 10, and 15 samples, respectively, for each scenario. 
To develop a keyword-specific model, we fine-tuned the pre-trained model for each keyword individually.

Each mini-batch consisted of 256 utterances, including 128 target, 96 non-target, and 32 noise samples.
Non-target samples were randomly chosen from 34 keywords, excluding the target, and both target and non-target samples underwent the same data augmentation process used during pre-training.
Noise samples were drawn from the background noise recordings in the GSC V2 dataset.
For fine-tuning, we used the AdamW optimizer with an initial learning rate of $10^{-5}$, which was halved every 20 epochs for a total of 150 epochs, with a weight decay of $10^{-5}$.

The official validation and test sets of GSC v2 were employed for model selection and evaluation.
Table \ref{tab:tab1} shows the average counts of positive and negative samples for the corresponding keywords.
To simulate real-world conditions, we generated noisy and reverberant speech by convolving synthetic room impulse responses (RIRs) from the OpenSLR dataset \cite{Ko-ICASSP} and adding noise from the MUSAN dataset \cite{Snyder-arxiv}, with signal-to-noise ratios (SNRs) ranging from 5 to 25 dB. 
We expanded the validation and test sets to be four times larger than their original sizes.
We evaluated model performance using two metrics: the Equal-Error-Rate (EER) \cite{Lee-INTERSPEECH, Nishu-INTERSPEECH} and Average Precision (AP) \cite{He-ICLR, RPL-Jung-INTERSPEECH, Hu-INTERSPEECH}, both of which are widely used in KWS.
To mitigate potential randomness, the fine-tuning dataset was sampled randomly five times under each condition, and the results were averaged.
The entire fine-tuning process took less than one hour on a V100 GPU, with the best epoch determined by the AP on the validation set.

\begin{table}[t!]
\caption{AP (\%) for FW-adapter and TE classifier (clf). `BN/SE Gx' is BN/SE adaptation in group x from Fig. \ref{fig:fig1}. `\# Params' is the number of tunable parameters.}
\label{tab:tab2}
\scriptsize
\centering
\setlength{\tabcolsep}{9.4pt}
\begin{tabular*}{\columnwidth}{c|c|c|c|c|c}
\hline
Method    & TE clf  & \# Params  & Seen  & Unseen  & Avg. \\ \hline
PT               & \checkmark            & -              &  77.46             & 66.58   & 72.02
      \\ \hline\hline
FT (full-shot)               &     - & 2.21 M                      &   96.71       &    93.99  & 95.35
     \\ 
FT (15-shot)               &     - & 2.21 M                              &   71.81       & 62.27  & 67.04  \\
FT clf (15-shot)              &     - & 0.5 K                      &   58.88        &    41.41  & 50.14
       \\ \hline\hline
BN G0                 &    \checkmark                              &     0.5 K            &   79.98  & 71.02   & 75.50  \\ 
BN G1                 &    \checkmark                               &     1.5 K           &  79.64 & 71.62    & 75.63

    \\ 
BN G2                 &     \checkmark                              &    1.5 K            &  80.59 &    72.66  & 76.63

    \\ 
BN G3                 &      \checkmark                            &       1.5 K           &  80.49    & 72.65   & 76.57

  \\ 
BN G5                 &      \checkmark                             &        3.1 K          &   78.16 &    67.87  & 73.02

  \\ 
BN G6                 &      \checkmark                            &       1.0 K           &   77.64  &  66.84   & 72.24

    \\ 
BN                 &     \checkmark                              &    9.0 K            &    82.04 &  76.76   & 79.40
  \\ \hline\hline
SE G1 \& BN             &     \checkmark                            &       41.8 K          &   84.94 &  83.30    & 84.12
  \\ 
SE G2 \& BN             &     \checkmark                            &       41.8 K          &  85.05 &  83.32   & 84.19

  \\ 
\textbf{SE G3 \& BN}             &     \checkmark                            &       41.8 K          &   \textbf{85.14} &  \textbf{83.52}   &  \textbf{84.33} \\ \hline\hline
SE \& BN             &     -                            &       107.8 K          &   63.02 &   65.62   & 64.32
  \\ 
SE \& BN             &     \checkmark                             &    107.3 K            &   82.87 & 77.49  & 80.18     \\ \hline
\end{tabular*}
\vspace{-0.2cm}
\end{table}

\subsection{Results}

Table \ref{tab:tab2} presents an ablation study that evaluates the effectiveness of the FW-adapter and the TE classifier.
For simplicity, the experiment focuses on a 15-shot scenario with five keywords from both seen and unseen keywords: 1) Seen keywords: `follow', `happy', `house', `one', and `seven'; 2) Unseen keywords: `cat', `dog', `eight', `nine', and `off'.
The table reports the average AP (\%) values for both seen and unseen keywords, as well as the average between them (`Avg.'). 
`PT' and `FT' refer to the pre-trained and fully fine-tuned models without the adapter, respectively.
`FT clf' represents the model where the network is frozen, and only an additional classifier (clf) is fine-tuned.
For `PT', the score is obtained from the cosine similarity between AE and TE, indicating the use of TE classifier.

Interestingly, `FT (15-shot)' performs worse (67.04\%) than `PT' (72.02\%), as 15-shot samples are insufficient for fine-tuning all the parameters.
However, when using all available samples for fine-tuning (FT (full-shot)), the model achieves remarkable performance (95.35\%), although this is impractical due to the excessive cost of data collection. 
Table \ref{tab:tab1} shows the average number of training samples.
Freezing the model and fine-tuning only an additional classifier (FT clf) yields poor performance (50.14\%).
Comparing `PT' and `FT clf', the TE classifier clearly boosts few-shot KWS performance.

\begin{table}[]
\caption{Performance of TCFM in terms of AP (\%).}
\vspace{-0.1cm}
\label{tab:tab3}
\footnotesize
\centering
\begin{tabular}{c|c|c|c|c|c}
\hline
Location & FW-adapter & \# Params & Seen & Unseen & Avg. \\ \hline
G0       &  \checkmark          &   44.9 K   &    56.01    &   41.82  & 48.92  \\ 
G1       &  \checkmark          &   69.5 K   &    43.86    &  34.22  & 39.04  \\ 
G2       &  \checkmark          &   69.5 K   &  50.22      &  40.94  & 45.58   \\ 
G3       &  \checkmark          &   69.5 K   &   62.76     &  46.77  & 54.77  \\ 
G4       &  \checkmark          &   44.9 K   &    86.34    &  84.02  & 85.18   \\ 
G5       &  \checkmark          &  41.9 K    &   85.93     & 83.98 & 84.96  \\ \hline\hline
G3, G4, G5       &  \checkmark          &   72.7 K   &    71.51    & 63.28   & 67.40  \\ 
\textbf{G4, G5}       &  \checkmark          &  45.0 K  & \textbf{88.45}  &   \textbf{85.98}     &   \textbf{87.22}   \\ \hline\hline
G4, G5       &  -          &  3.2 K   &    79.94    &   68.29 &  74.12  \\ \hline
\end{tabular}
\vspace{-0.1cm}
\end{table}

We evaluate performance by individually or collectively adapting the BN layers in each group. Hereafter, we omit mentioning `(15-shot)' in the method, but all methods continue to use 15-shot samples.
Regardless of the adaptation location within the ECAPA-TDNN, BN adaptation consistently outperforms the pre-trained model for both seen and unseen keywords, validating our hypothesis that adjusting feature weights at each layer enables successful adaptation to a target keyword. 
Adopting BN adaptation across all groups (`BN') yields better results (79.40\%) than fine-tuning individual groups, indicating that the gains from individual group adaptations are complementary and cumulative.
Combining SE and BN adaptations yields further improvements.
However, contrary to BN, applying SE adaptation to all groups (`SE \& BN') performs worse compared to applying it individually to each group.
We hypothesize that this could be attributed to the excessive number of additional parameters required for few-shot KWS.
The best performance is an AP of 84.33\%, achieved with `SE G3 \& BN', without adding any extra parameters.

Table \ref{tab:tab3} presents the ablation results on TCFM, where `FW-adapter' corresponds to `SE G3 \& BN'.
Conditioning keyword information at the lower layers (i.e., G0 to G3) degrades performance. 
We suspect that directly modifying their features with limited samples impairs performance, as these early layers generate fundamental features that eventually form keyword-specific representations.
In contrast, the FW-adapter merely adjusts feature aggregation, enabling effective adaptation at any layer and improving performance compared to the pre-trained model.
Our results suggest that applying TCFM to higher layers is suitable for conditioning TE, consistent with AdaKWS \cite{Navon-ICASSP}, where KAMs are inserted just before the final classifier.
Optimal performance is achieved when applying TCFM to G4 and G5, boosting the AP of TF-KWS (i.e., `PT' in Table \ref{tab:tab2}) from 72.02\% to 87.22\%.
Although TCFM slightly increases the total number of parameters (by 3.2K, representing 0.14\% of the original count), the performance gain is substantial.

To better understand how TCFM adapts based on the conditioning vector, Fig. \ref{fig:fig3} visualizes the outputs of the trained LAFs from G4 and G5, conditioned on TEs extracted from six keywords.
To account for varying activation ranges, the plots display LAFs subtracted by the average of basic activations: 
\begin{equation}\label{eqn:eq3}
 y = \textrm{LAF}(h|\textrm{TE}_k) - a^{-1} \sum_{i=1}^a A_i(h),
\end{equation}
where $h \in [-3,3]$ and $k$ is the selected keyword.
It is evident that different keywords exhibit distinct LAF patterns, and each keyword's LAF varies across layers. 
As a result, it can be observed that TCFM conditions the keyword information of TE by learning a weighted combination of basic activation functions based on TE.

Finally, Table \ref{tab:tab4} compares the performance of the TA-adapter with other baseline approaches using all 35 keywords. `TA-adapter' corresponds to the best-performing model from Table \ref{tab:tab3}.
We report the average values of Seen and Unseen for both AP and EER.
Here, `AdaMS' and `RPL' are TF-KWS models trained on the same out-of-domain dataset \cite{Speechocean-DB} without fine-tuning.
Notably, `RPL' corresponds to our pre-trained model.
The remaining models are few-shot KWS baselines.
For a fair comparison, all few-shot KWS models share the same experimental setup, including datasets, model architectures, and pre-trained model, except for their training strategies.
`2-class clf' and `3-class clf' are few-shot learning methods for speech-enrolled flexible KWS without freezing the pre-trained model. `2-class clf' utilizes binary classification to predict the probability of the keyword’s presence. `3-class clf' classifies target, non-target, and background noise categories.
`AdaKWS' replaces the TA-adapter with KAMs, with the rest of the architecture unchanged, allowing for a comparison between KAM-based and TA-adapter methods.
Specifically, KAMs are inserted after the pre-trained acoustic encoder, and the entire model is fine-tuned in the same manner as our approach.
Across all scenarios, the TA-adatper consistently outperforms the baseline systems, including TF-KWS models (`AdaMS' and `RPL') and few-shot learning approaches (`2-class clf', `3-class clf', and `AdaKWS').
Notably, the performance gap widens under lower resource conditions.

\begin{figure}[t]
\centering
     \begin{subfigure}[b]{0.239\textwidth}
         \centering
         \includegraphics[width=\textwidth]{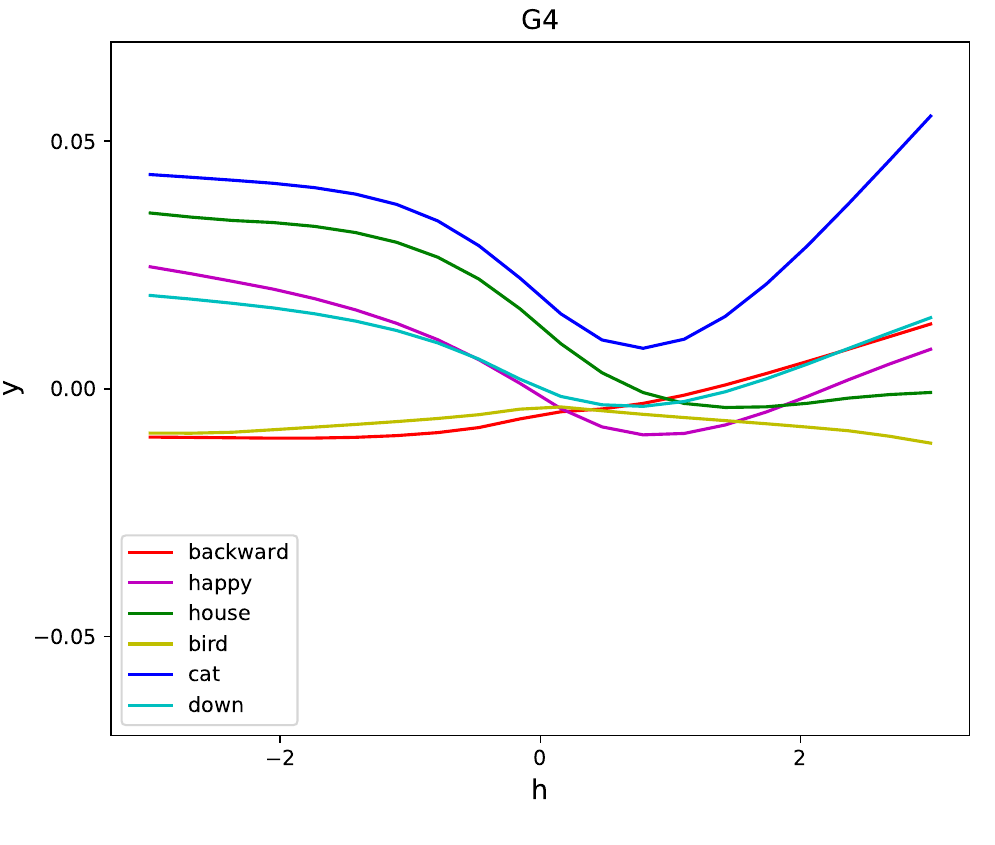}
         \label{fig:fig3_a}
     \end{subfigure}
     \begin{subfigure}[b]{0.239\textwidth}
         \centering
         \includegraphics[width=\textwidth]{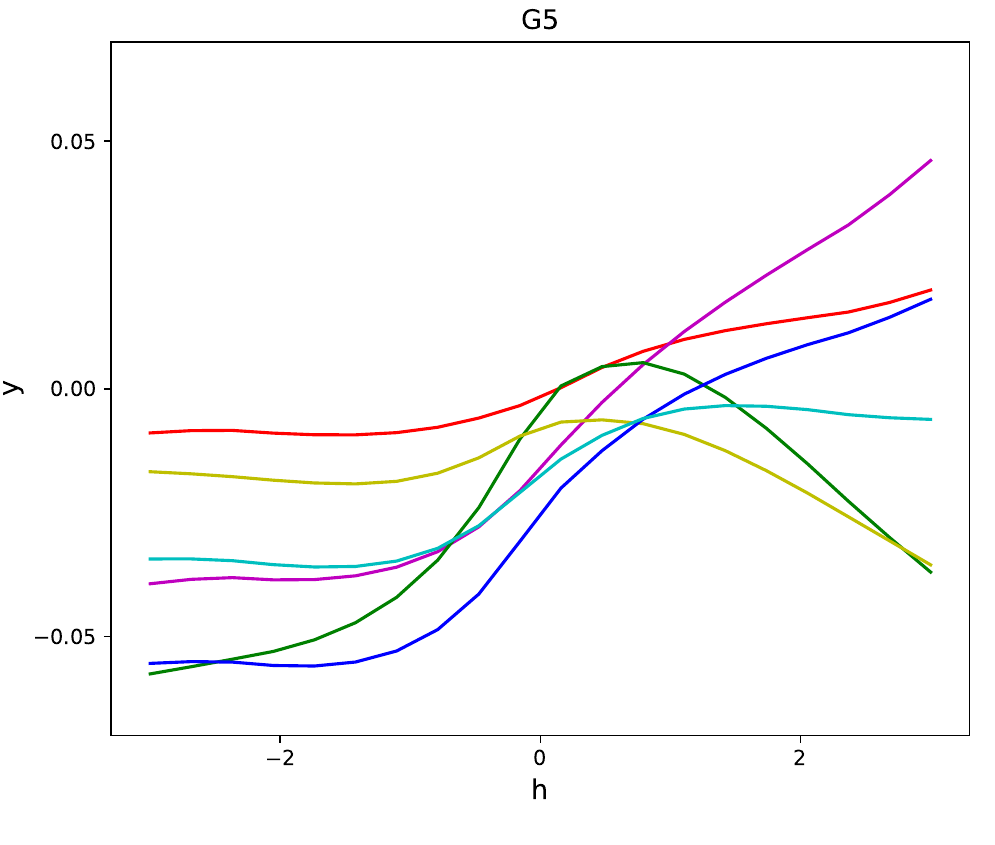}
         \label{fig:fig3_b}
     \end{subfigure}
\vspace{-0.9cm}
\caption{Plot of normalized outputs of trained LAFs from G4 and G5 conditioned on TEs extracted from six keywords (`backward', `happy', `house', `bird', `cat', and `down'). The plots emphasize that LAF exhibit varying profiles across different keywords and layers.}
\label{fig:fig3}
\end{figure}

\begin{table}[t]
\footnotesize
\centering
\caption{Performance comparison with baseline approaches in terms of AP (\%) and EER (\%). \textdagger: results without fine-tuning.}
\vspace{-0.10cm}
\label{tab:tab4}
\begin{tabular}{c|cc|cc|cc}
\hline
\multirow{2}{*}{Method} & \multicolumn{2}{c|}{5-shot} & \multicolumn{2}{c|}{10-shot} & \multicolumn{2}{c}{15-shot} \\ \cline{2-7} 
    & \multicolumn{1}{c|}{AP}  & EER & \multicolumn{1}{c|}{AP} & EER & \multicolumn{1}{c|}{AP} & EER \\ \hline
AdaMS \cite{RPL-Jung-INTERSPEECH} & \multicolumn{1}{c|}{74.44\textsuperscript{\textdagger}} & 8.25\textsuperscript{\textdagger} & \multicolumn{4}{c}{\multirow{2}{*}{N/A}}   \\ 
RPL \cite{Jung-INTERSPEECH-AdaMS} & \multicolumn{1}{c|}{77.93\textsuperscript{\textdagger}} & 7.94\textsuperscript{\textdagger} & \multicolumn{4}{c}{}   \\ \hline\hline
2-class clf \cite{Yuan-INTERSPEECH} & \multicolumn{1}{c|}{50.97} & 20.45 & \multicolumn{1}{c|}{65.53}  & 15.27  & \multicolumn{1}{c|}{72.38}  & 10.49   \\ 
3-class clf \cite{Mazumder-INTERSPEECH} & \multicolumn{1}{c|}{52.54} & 18.91 & \multicolumn{1}{c|}{68.67}  & 13.22   & \multicolumn{1}{c|}{73.98}  & 9.87   \\ 
AdaKWS \cite{Navon-ICASSP} & \multicolumn{1}{c|}{58.32} & 14.02 & \multicolumn{1}{c|}{59.13}  & 13.33   & \multicolumn{1}{c|}{61.67}  & 12.59   \\\hline\hline
\textbf{TA-adapter} & \multicolumn{1}{c|}{87.63} & 5.09 & \multicolumn{1}{c|}{89.15}  & 4.66   & \multicolumn{1}{c|}{90.38}  & 4.56   \\ \hline
\end{tabular}
\vspace{-0.15cm}
\end{table}

\section{Conclusion}
\label{sec:majhead}

This paper introduces the TA-adapter for addressing the few-shot transfer learning problem in TF-KWS.
The TA-adapter utilizes the text embedding to condition keyword information into the acoustic encoder and to generate a keyword score.
Also, BN layers and SE modules are adapted using only a few samples of the target keyword.
Experimental results demonstrate that the TA-adapter effectively overcomes the challenges of few-shot KWS.
Specifically, the TA-adapter boosts the pre-trained model's average precision from 77.93\% to 87.63\% using just 5 target samples.
Since the TA-adapter only modifies a small subset of the model, it enables seamless reversion to the original pre-trained model.
In future work, we plan to leverage text-to-speech (TTS) technology to generate synthetic data, enabling zero-shot transfer learning for TF-KWS.

\clearpage
\bibliographystyle{IEEEtran}
\bibliography{template}

\begin{thebibliography}{10}
\providecommand{\url}[1]{#1}
\csname url@samestyle\endcsname
\providecommand{\newblock}{\relax}
\providecommand{\bibinfo}[2]{#2}
\providecommand{\BIBentrySTDinterwordspacing}{\spaceskip=0pt\relax}
\providecommand{\BIBentryALTinterwordstretchfactor}{4}
\providecommand{\BIBentryALTinterwordspacing}{\spaceskip=\fontdimen2\font plus
\BIBentryALTinterwordstretchfactor\fontdimen3\font minus \fontdimen4\font\relax}
\providecommand{\BIBforeignlanguage}[2]{{%
\expandafter\ifx\csname l@#1\endcsname\relax
\typeout{** WARNING: IEEEtran.bst: No hyphenation pattern has been}%
\typeout{** loaded for the language `#1'. Using the pattern for}%
\typeout{** the default language instead.}%
\else
\language=\csname l@#1\endcsname
\fi
#2}}
\providecommand{\BIBdecl}{\relax}
\BIBdecl

\bibitem{Chen14-ICASSP}
G.~Chen, C.~Parada, and G.~Heigold, ``Small-footprint keyword spotting using deep neural networks,'' in \emph{Proc. IEEE International Conference on Acoustics, Speech and Signal Processing (ICASSP)}, 2014, pp. 4087--4091.

\bibitem{Sainath-INTERSPEECH}
T.~N. Sainath and C.~Parada, ``Convolutional neural networks for small-footprint keyword spotting,'' in \emph{Proc. Interspeech}, 2015, pp. 1478--1482.

\bibitem{TANG17-ICASSP}
R.~Tang and J.~Lin, ``Deep residual learning for small-footprint keyword spotting,'' in \emph{Proc. IEEE International Conference on Acoustics, Speech and Signal Processing (ICASSP)}, 2017, pp. 5484--5488.

\bibitem{Chen-ICASSP}
G.~Chen, C.~Parada, and T.~N. Sainath, ``Query-by-example keyword spotting using long short-term memory networks,'' in \emph{Proc. IEEE International Conference on Acoustics, Speech and Signal Processing (ICASSP)}, 2015, pp. 5236--5240.

\bibitem{Huang-ICASSP}
J.~Huang, W.~Gharbieh, H.~S. Shim, and E.~Kim, ``Query-by-example keyword spotting system using multi-head attention and soft-triple loss,'' in \emph{Proc. IEEE International Conference on Acoustics, Speech and Signal Processing (ICASSP)}, 2021, pp. 6858--6862.

\bibitem{Kurmi-INTERSPEECH}
K.~R, V.~K. Kurmi, V.~Namboodiri, and C.~V. Jawahar, ``Generalized keyword spotting using {ASR} embeddings,'' in \emph{Proc. Interspeech}, 2022, pp. 126--130.

\bibitem{He-ICLR}
W.~He, W.~Wang, and K.~Livescu, ``Multi-view recurrent neural acoustic word embeddings,'' in \emph{Proc. International Conference on Learning Representations (ICLR)}, 2017.

\bibitem{Jung-INTERSPEECH-AdaMS}
M.~Jung and H.~Kim, ``{AdaMS}: Deep metric learning with adaptive margin and adaptive scale for acoustic word discrimination,'' in \emph{Proc. Interspeech}, 2023, pp. 3924--3928.

\bibitem{Nishu-INTERSPEECH}
K.~Nishu, M.~Cho, and D.~Naik, ``Matching latent encoding for audio-text based keyword spotting,'' in \emph{Proc. Interspeech}, 2023, pp. 1613--1617.

\bibitem{Lee-INTERSPEECH}
Y.-H. Lee and N.~Cho, ``{PhonMatchNet}: Phoneme-guided zero-shot keyword spotting for user-defined keywords,'' in \emph{Proc. Interspeech}, 2023, pp. 3964--3968.

\bibitem{RPL-Jung-INTERSPEECH}
Y.~Jung, S.~Lee, J.-Y. Yang, J.~Roh, C.~W. Han, and H.-Y. Cho, ``Relational proxy loss for audio-text based keyword spotting,'' in \emph{Proc. Interspeech}, 2024, pp. 327--331.

\bibitem{Wang-CVPR}
X.~Wang, X.~Han, W.~Huang, D.~Dong, and M.~R. Scott, ``Multi-similarity loss with general pair weighting for deep metric learning,'' in \emph{Proc. the IEEE/CVF Conference on Computer Vision and Pattern Recognition (CVPR)}, 2019, pp. 5022--5030.

\bibitem{Bluche-INTERSPEECH}
T.~Bluche and T.~Gisselbrecht, ``Predicting detection filters for small footprint open-vocabulary keyword spotting,'' in \emph{Proc. Interspeech}, 2020, pp. 2552--2556.

\bibitem{Kao-SLT}
W.-T. Kao, Y.-K. Wu, C.-P. Chen, Z.-S. Chen, Y.-P. Tsai, and H.-Y. Lee, ``On the efficiency of integrating self-supervised learning and meta-learning for user-defined few-shot keyword spotting,'' in \emph{Proc. SLT}, 2022, pp. 414--421.

\bibitem{speechcommandsv2}
\BIBentryALTinterwordspacing
P.~{Warden}, ``Speech {C}ommands: A dataset for limited-vocabulary speech recognition,'' \emph{arXiv:1804.03209}, 2018. [Online]. Available: \url{https://arxiv.org/abs/1804.03209}
\BIBentrySTDinterwordspacing

\bibitem{Mazumder-INTERSPEECH}
M.~Mazumder, C.~Banbury, J.~Meyer, P.~Warden, and V.~J. Reddi, ``Few-shot keyword spotting in any language,'' in \emph{Proc. Interspeech}, 2021, pp. 4214--4218.

\bibitem{Awasthi-INTERSPEECH}
A.~Awasthi, K.~Kilgour, and H.~Rom, ``Teaching keyword spotters to spot new keywords with limited examples,'' in \emph{Proc. Interspeech}, 2021, pp. 4254--4258.

\bibitem{Jung-metric-ICASSP}
J.~Jung, Y.~Kim, J.~Park, Y.~Lim, B.-Y. Kim, Y.~Jang, and J.~S. Chung, ``Metric learning for user-defined keyword spotting,'' in \emph{Proc. IEEE International Conference on Acoustics, Speech and Signal Processing (ICASSP)}, 2023, pp. 1--5.

\bibitem{Yuan-INTERSPEECH}
J.~Yuan, Y.~Shi, L.~Li, D.~Wang, and A.~Hamdulla, ``Few-shot keyword spotting from mixed speech,'' in \emph{Proc. Interspeech}, 2024, pp. 5063--5067.

\bibitem{Rebuffi-NIPS}
S.-A. Rebuffi, H.~Bilen, and A.~Vedaldi, ``Learning multiple visual domains with residual adapters,'' in \emph{Proc. Advances in Neural Information Processing Systems (NIPS)}, 2017.

\bibitem{Pfeiffer-EMNLP}
J.~Pfeiffer, A.~Rücklé, C.~Poth, A.~Kamath, I.~Vulić, S.~Ruder, K.~Cho, and I.~Gurevych, ``Adapterhub: A framework for adapting transformers,'' in \emph{Proc. EMNLP}, 2020.

\bibitem{Navon-ICASSP}
A.~Navon, A.~Shamsian, N.~Glazer, G.~Hetz, and J.~Keshet, ``Open-vocabulary keyword-spotting with adaptive instance normalization,'' in \emph{Proc. IEEE International Conference on Acoustics, Speech and Signal Processing (ICASSP)}, 2024, pp. 11\,656--11\,660.

\bibitem{Huang-ICCV}
X.~Huang and S.~J. Belongie, ``Arbitrary style transfer in real-time with adaptive instance normalization,'' in \emph{Proc. IEEE International Conference on Computer Vision (ICCV)}, 2017, pp. 1510--1519.

\bibitem{Ramos-ICLR}
A.~G. C.~P. Ramos, A.~Mehrotra, N.~D. Lane, and S.~Bhattacharya, ``Conditioning sequence-to-sequence networks with learned activations,'' in \emph{Proc. International Conference on Learning Representations (ICLR)}, 2022.

\bibitem{Sarfjoo-INTERSPEECH}
S.~S. Sarfjoo, S.~R. Madikeri, P.~Motlicek, and S.~Marcel, ``Supervised domain adaptation for text-independent speaker verification using limited data,'' in \emph{Proc. Interspeech}, 2020, pp. 3815--3819.

\bibitem{Wang-INTERSPEECH}
T.~Wang, L.~Li, and D.~Wang, ``{SE}/{BN} adapter: Parametric efficient domain adaptation for speaker recognition,'' in \emph{Proc. Interspeech}, 2024, pp. 2145--2149.

\bibitem{Hu-CVPR}
J.~Hu, L.~Shen, and G.~Sun, ``Squeeze-and-excitation networks,'' in \emph{Proc. the IEEE/CVF Conference on Computer Vision and Pattern Recognition (CVPR)}, 2018, pp. 7132--7141.

\bibitem{Desplanques-arxiv}
B.~Desplanques, J.~Thienpondt, and K.~Demuynck, ``{ECAPA-TDNN}: Emphasized channel attention, propagation and aggregation in {TDNN} based speaker verification,'' \emph{arXiv:2005.07143}, 2020.

\bibitem{Zhao-ICASSP}
Z.~Zhao, Z.~Li, W.~Wang, and P.~Zhang, ``{PCF}: {ECAPA-TDNN} with progressive channel fusion for speaker verification,'' in \emph{Proc. IEEE International Conference on Acoustics, Speech and Signal Processing (ICASSP)}, 2023, pp. 1--5.

\bibitem{Liao-ICASSP}
C.~L. et~al., ``Dynamic {TF-TDNN}: Dynamic time delay neural network based on temporal-frequency attention for dialect recognition,'' in \emph{Proc. IEEE International Conference on Acoustics, Speech and Signal Processing (ICASSP)}, 2023, pp. 1--5.

\bibitem{Li-INTERSPEECH}
H.~Li, B.~Yang, Y.~Xi, L.~Yu, T.~Tan, H.~Li, and K.~Yu, ``Text-aware speech separation for multi-talker keyword spotting,'' in \emph{Proc. Interspeech}, 2024, pp. 337--341.

\bibitem{Gao-TPAMI}
S.~Gao, M.-M. Cheng, K.~Zhao, X.~Zhang, M.-H. Yang, and P.~H.~S. Torr, ``Res2net: A new multi-scale backbone architecture,'' \emph{IEEE TPAMI}, 2019.

\bibitem{Speechocean-DB}
\BIBentryALTinterwordspacing
DataOceanAI, ``King-{ASR}-066,'' 2015. [Online]. Available: \url{https://en.speechocean.com/datacenter/details/1446.html}
\BIBentrySTDinterwordspacing

\bibitem{Ko-ICASSP}
T.~Ko, V.~Peddinti, D.~Povey, M.~L. Seltzer, and S.~Khudanpur, ``A study on data augmentation of reverberant speech for robust speech recognition,'' in \emph{Proc. IEEE International Conference on Acoustics, Speech and Signal Processing (ICASSP)}, 2017, pp. 5220--5224.

\bibitem{Snyder-arxiv}
D.~Snyder, G.~Chen, and D.~Povey, ``Musan: A music, speech, and noise corpus,'' \emph{arXiv:1510.08484}, 2015.

\bibitem{Hu-INTERSPEECH}
Y.~Hu, S.~Settle, and K.~Livescu, ``Multilingual jointly trained acoustic and written word embeddings,'' in \emph{Proc. Interspeech}, 2020, pp. 1052--1056.

\end{thebibliography}

\end{document}